\documentclass[aps,preprint,showpacs]{revtex4}
\usepackage{amsmath,epsfig}
\usepackage{graphicx}
\begin{document}
\title{Many-polaron system in Fractional Dimensional space within 
the Plasmon Pole Approximation} 
\author{K. M. Mohapatra, B. K. Panda, S. Kar}
\date{\today}
\begin{abstract}
The polaron binding energy and effective mass in a degenerate polar
gas is calculated in the fractional-dimensional approach under plasmon
pole approximation.The effect of carrier densities on the static and dynamic 
screening correction of the electron-phonon interaction and electron-electron interaction to the polaronic propertis is calculated from electron 
self-energies within the second-order perturbation method. The Hubbard local 
field factor has been used for the static screening correction in the 
polaronic properties. We found that polaronic properties decrease with 
increase with carrier density and dimensionality of the system.
\end{abstract}
\maketitle
\section{Introduction}
When an electron in the bottom of the conduction band of a polar semiconductor 
moves, its Coulomb field displaces the positive and negative ions with 
respect to each other producing polarization field. The electron 
with the associated phonon cloud is known as a polaron. 

The density oscillation in a medium creates plasmons. The interaction  
between an electron and plasmons can be described by Fr\"olich Hamiltonian. 
The interaction of the electron with plasmon cloud is termed as the 
plasmapolaron. The binding energy is shifted and electron mass in enhanced.

The Fr\"ohlich polaron results when the interaction between 
electron and longitudinal optic 
(LO) phonon is described in the Fr\"ohlich form characterized by the 
dimensionless coupling constant $\alpha$. The polaron in an undoped polar
material calculated with the unscreened electron-phonon interaction is 
termed as a single-polaron. 

In the weak coupling limit $(\alpha<1)$,
the perturbation theory is applied to calculate ground state energies and 
effective masses of the single-polaron in three-dimensional 
(3D) polar materials\cite{Frolich}, two-dimensional (2D) quantum well (QW) 
structures\cite{Mason} and one-dimensional
(1D) quantum wire structures\cite{Peeters}. The polaron properties 
were also calculated in the multidimensional space $(nD)$ where the space has 
an integer dimension $n$. In the $nD$ space, the expressions for the 
ground state energy  and effective mass 
were correctly expressed in terms of $n$ in the weak\cite{Serenius},
intermediate\cite{Ashoka1} and strong coupling\cite{Ashoka2} limits. 
Since the Coulomb interaction in Fourier method is divergent in the 
1D systems, the multidimensional method fails to calculate polaron properties 
in the quantum wire structures. Therefore, the multidimensional space method 
finds the polaron properties in 2D and 3D systems quite accurately. 
When the width of the QW is extremely narrow and 
the barrier potential that causes
the in-plane confinement is infinite, the system is purely 2D. The dimension 
increses monotonically with the increase of the well width and the infinitely
wide well exhibits the three-dimensional (3D) behaviour.
In a finite QW with narrow well width, the electron envelope function
spreads into the barrier region partially restoring  the 3D
characteristics of the system. The properties of the QW are determined by the 
parameters of the barrier materials. In the QW with large well width, 
the properties of the polaron are calculated taking the bulk values of the 
well material. This has been demonstrated in the
calculation of exciton binding energy\cite{Jai} and polaron properties 
as a function of well width\cite{Matos}.
Consequently, the QW with finite well width and barrier height shows
fractional dimensional behaviour $(\beta D)$ where $2<\beta<3$. 
The anisotropic interactions in an anisotropic solid
are treated as ones in an isotropic
fractional dimensional space, where the dimension is
determined by the degree of anisotropy.
Thus only a single parameter known as the degree of
dimensionality $\beta$ is needed to describe the system.
The fractional dimensional space is not a vector space and
the coordinates in this space are termed as
{\em pseudocoordinates}\cite{Stillinger}.

  The fractional dimensional space method has been applied to study exciton,
biexciton, magnetoexciton and impurity states. The single-polaron binding 
energy and effective mass in the weak coupling limit 
have been derived in rectangular and parabolic QWs. The results are found 
to increase with decreasing confinemnt.  

When the carrier density is high, the electron-phonon
interaction is dynamically screened by the frequency dependent 
dielectric function and 
the resulting polaron is termed as the 
many-polaron.  However, for systems with much larger plasma frequency than the 
LO phonon frequency, the static dielectric function can also be a good 
approximation. This happens when the carrier density is very high in the 
system. The properties of many-polaron in the doped ZnS was studied 
by da Costa and Studart\cite{Studart} including exchange-correlation 
effects beyond RPA and found that the ground state energies and effective masses
are influenced by the exchange-correlation effects.  

The static properties such as the binding energy and optical conductivity
have been studied by including the dynamical scrrening method.
The effective mass of the polaron has been found from the 
optical conductivity. The results are markedly different from those 
calculated using the static dielectric function\cite{Tempere1,Tempere2,Hameeuw}.  

In the present work we study the many-polaron 
system by the static and dynamically screened electron-phonon interaction and electron-electron interaction.

\section{Polaron Binding Energy and effective mass due to static 
screening effect}

In the second-order  perturbation method, the polaron binding energy 
is calculated from the electron self-energy due to 
electron-phonon interaction as\cite{Mahan}
\begin{equation}
E_{\beta D}=-\Sigma_{\beta D}({\bf k},\xi_{k})|_{{\bf k}=0},
\end{equation}
where $\xi_{k}$ is the electron energy with parabolic band dispersion 
$(\hbar^2k^2/2m_{b})$.  
The effective mass in the same method is defined as
\begin{equation}
\frac{m_{b}}{m^{\ast}}=1+\biggl(\frac{m_{b}}{\hbar^2}\biggr)
\frac{\partial^{2}\Sigma_{\beta D}({\bf k},\xi_{k})}{\partial k^2}\Biggr|_{{\bf k}=0}
\end{equation}
where $m_{b}$ is the band electron mass.  

The leading Feynman diagram contribution to the elctron self energy due
the electron-phonon interaction is given by
$$\Sigma_d(p,\iota k_m)=-\frac{1}{\beta}\sum_{m,q}
V_{eff}(q,\omega)G^0(\iota\omega_n,\xi_{p+q})$$

Here 
$$V_{eff}(q,\omega)=\frac{M_{d}(q,\omega)^2}{\epsilon_{d}(q,\omega)^2}
D_{d}(q,\omega)$$ 
is the effective potential for electron-phonon interaction.

 $D_d(q,\omega)$ is the renormalizesd phonon propagator given by 

$$D_{d}(q,\omega)=\frac{D_0(q,\omega)}{1-|M_d(q)|^2 D_0(q,\omega)
\epsilon_{d}(q,\omega)}$$

$$D_0(q,\omega)=\frac{2\omega_{LO}}{\omega^2-\omega_{LO}^2}$$. 
is the unpertured phonon propagaotor.

The free electron Green function $G^{0}(\iota\omega_{n},\xi_{(p+q)})$
is given as
$$G^{0}(\iota\omega_{n},\xi_{(p+q)})=\frac{1}
{\iota p_{m}+i\hbar\omega_{n}-\xi_{p+q}}$$
Using eq(), eq() in eq() we get the self energy equation becomes 

$$\Sigma({\bf k},ik_{m})=-\frac{1}{\beta}\sum_{{\bf q}}\frac{|M(q)|^{2}}
{\epsilon(q,rs)^2}\sum^{\infty}_{i\omega_{n}=-\infty}
\frac{2\omega_{LO}}{(\iota \omega_n)^2-(\omega_{LO})^2}\times
\frac{1}{\iota p_{m}+i\hbar\omega_{n}-\xi_{p+q}}$$
Using Matsubara frequency summation the electron-phonon self energy becomes

$$\Sigma({\bf k},ik_{m})=-\frac{1}{\beta}\sum_{{\bf q}}\frac{|M(q)|^{2}}
{\epsilon(q,rs)^2}\times\Biggr[\frac{1+n_B(\omega_{LO}-n_F{\xi_{p+q}}}
{\iota p_{m}-\hbar\omega_{LO}-\xi_{p+q}}+\frac{n_B(\omega_{LO}+n_F{\xi_{p+q}}}
{\iota p_{m}+\hbar\omega_{LO}-\xi_{p+q}}\Biggr]$$
At zero temperature both $n_F$ and $n_B$ vanish and the self energy 
equation becomes

$$\Sigma({\bf k},ik_{m})=-\frac{1}{\beta}\sum_{{\bf q}}\frac{|M(q)|^{2}}
{\epsilon(q,rs)^2}
\times\Biggr[\frac{1}{\xi_p-\xi_{p+q}-\hbar\omega_{LO}}\Biggr]$$

The electron-phonon interaction term is expressed as 
\begin{equation}
M_{\beta D}(q)=-i\hbar\omega_{LO}\Biggl(\frac{(4\pi)^{\frac{\beta-1}{2}}\Gamma\biggl(\frac{\beta-1}{2}\biggr)R_{p}\alpha}
{q^{\beta-1}\Omega_{\beta }}\Biggr)^{\frac{1}{2}},
\end{equation}
where $\Gamma$ is the Euler-gamma function and
$R_p=\sqrt{\hbar/2m_b\omega_{LO}}$ is the polaron radius. 
The dimensionless coupling constant $\alpha$ is defined as
\begin{equation}
\alpha=\frac{e^2}{2\hbar\omega_{LO}R_{p}}
\biggl(\frac{1}{\epsilon_{\infty}}-\frac{1}{\epsilon_{0}}\biggr),
\end{equation}
where $\epsilon_{0}$ and $\epsilon_{\infty}$ are the static and high-frequency dielectric constants, respectively.  
The static dielectric function including the local-field factor is defined as
\begin{equation}
\epsilon_{\beta D}(q,rs)=\epsilon_{\infty}\Biggl[\frac{1-[1-G_{\beta D}(q,rs)]
V_{\beta D}(q)\chi_{\beta D}(q,rs)}
{1+G_{\beta D}(q,rs)V_{\beta }(q)\chi_{\beta D}(q,rs)}\Biggr],
\end{equation}
where $G_{\beta D}(q,rs)$ is the Hubbard local-field-factor given by

\begin{equation}
G_{\beta D}(q,rs)=\frac{1}{2}\frac{q^{\beta-1}}{(q^2+k^{2}_{F})^{\frac{\beta-1}{2}}}.
\end{equation}
The dimensionless density parameter $rs$ is given by
\begin{equation}
k_{F}r_{s}a_{B}=\biggl[2^{\beta-1}\Gamma^{2}
\biggl(1+\frac{\beta}{2}\biggr)\biggr]^{\frac{1}{\beta}}
\end{equation}
 where $a_{B}$ is Bohr atomic radius and $k_{F}$ is Fermi wave vector.

The irreducible polarizability function 
 $\chi_{\beta D}(q,rs)$ is defined as
\begin{equation}
\chi^{0}_{\beta D}(q,r_{s})=\sum_{{\bf k}}\frac{n_{F}({\bf k}+{\bf q})-
n_{F}({\bf k})}{\xi_{{\bf k}+{\bf q}}-\xi_{\bf k}}.
\end{equation}
In the fractional dimensional method the sum over $q$  is 
transferred to integration as 

$$\sum_{\bf q}\cdot\cdot\cdot=\frac{V}{(2\pi)^d}\frac{2\pi^{\frac{d-1}{2}}}
{\Gamma\small(\frac{d-1}{2}\small)}\int dq q^{d-1}\int^{\pi}_{0}
(\sin\theta)^{d-2}\cdot\cdot\cdot\label{eq:sum}$$

Using Eq.(10) in Eq. (9), we find
\begin{equation}
\chi^{0}_{\beta D}(q,rs)=-\frac{2^{3-\beta}m_{b}k^{\beta}_{F}}
{\pi^{\frac{\beta-1}{2}}\hbar^2\beta q^{2}\Gamma\biggl(\frac{\beta-1}{2}\biggr)} \int^{\pi}_{0}{_2F_1}\biggl(1,\frac{\beta}{2};
\frac{\beta+2}{2};\frac{4 k^{2}_{F}\cos^{2}\theta}{q^2}\biggr)\sin^{\beta-2}\theta d\theta,
\end{equation}
where $_{2}F_{1}$ is the Gauss hypergeometric function.

The Fourier transform of $e^2/r$ in the fractional dimensional space 
is obtained as
\begin{equation}
V_{\beta D}(q)=\frac{(4\pi)^{\frac{\beta-1}{2}}e^2
\Gamma\biggl(\frac{\beta-1}{2}\biggr)}
{q^{\beta-1}}\label{eq:Coul}
\end{equation}

Using Eqs. (6), (7), (10) and (12) in Eq. (1), the binding energy is
obtained as
\begin{equation}
E_{\beta D}=-\alpha\hbar\omega_{LO}\frac{\Gamma\biggl(\frac{\beta}{2}\biggr)}
{\sqrt{\pi}\Gamma\biggl(\frac{\beta}{2}\biggr)}
\int^{\infty}_{0}\frac{dq}{\epsilon_{\beta}(q,r_{s})(q^2+1)}.
\end{equation}

 Similarly the effective mass in Eq.(2) can be obtained as
\begin{equation}
\frac{m_{b}}{m^{\ast}}=1-4\alpha
\frac{\Gamma\biggl(\frac{\beta-1}{2}\biggr)}
{\sqrt{\pi}\beta\Gamma\biggl(\frac{\beta}{2}\biggr)}
\int^{\infty}_{0}\frac{q^2dq}{\epsilon^{2}_{\beta D}(q,rs)(q^2+1)^{3}}.
\end{equation}

For nondegenerate systems$(rs\rightarrow \infty)$, $\epsilon_{\beta D}=1$. The integrals in above Eqs(11) and (12) can be analytically evaluated.Now
The binding energy s derived as
\begin{equation}
E_{\beta D}=-\frac{1}{2}\alpha\hbar\omega_{LO}
\frac{\sqrt{\pi}\Gamma\biggl(\frac{\beta-1}{2}\biggr)}
{\Gamma\biggl(\frac{\beta}{2}\biggr)}
\end{equation}

and the effective mass is given by
\begin{equation}
\frac{m_b}{m^{\ast}}=1+\frac{1}{4}\alpha
\frac{\sqrt{\pi}\Gamma\biggl(\frac{\beta-1}{2}\biggr)}
{\beta\Gamma\biggl(\frac{\beta}{2}\biggr)},
\end{equation}
\section{result and discussion}
we have taken the several parameters of GaAs to calculate 
polaron properties.The band mass  $m_b$=0.067$m_{0}$, where $m_{0}$ 
is the electron mass,   
$\epsilon_{\infty}=13.18$ and $\epsilon_{0}=10.89$. The value of 
$\alpha$=0.03 which is appropriate for calculating polaron properties 
in the weak coupling limit. The LO phonon energy $(\hbar\omega_{LO})$ 
is taken as 36.25 meV.

The polaron binding energy and effective masses due to static  
screening correction of electron-phonon interaction
is calculated for several $r_{s}$ values 
for dimensions $\beta$=2, 2.5 and 3 are shown in Table I.
Both binding  energies and effective masses are found to increase with 
the increasing  $r_{s}$. This suggests that the polaron properties 
decrease with the  increasing carrier density. 
This results due to screening of the electron-phonon interaction. 

The polaron properties also decrease as the dimension decreases. As the 
dimension of the system decreases, the system becomes more confined. 
The confinement of the system decreases the physical properties. 

Although the static Hubbard's local-field-factor for the screening of the 
electron-phonon interaction correctly predicts  
$r_{s}$ and $\beta$ dependence of the physical properties, it does not  
include the exchange and correlation effects properly. It is much 
higher than the dynamic local field factor as the later correctly includes  
the exchange and correlation effects. Therefore it is required that 
we calculate the dynamic local field factor using quantum version of the 
STLS method and screen the electron-phonon interaction term. 
Such a work is in progress in our group.

\section{Polaron Binding Energy and effective mass due to dynamic 
	screening effect}
Conventionally the Raleigh-Schr\"odinger (RS) and Tamm-Dancoff approximation 
to the Brilloun-Wigner (TD-BW)
perturbation theories are employed to calculate ground state energies and 
effective masses of the single-polaron state\cite{Mahan}. The effective masses 
calculated in the RS perturbation theory were more consistent than the 
TD-BW methods. This scenario did not change when the effective masses 
of the many-polaron system were calculated in bulk semiconductors. 
We have therefore 
taken the RS perturbation theory to calculate binding energies and 
effective masses of fractional dimensional many-polaron.  

In the RS perturbation method, the polaron binding energy is calculated from the 
electron self-energy due to electron-phonon interaction as
\begin{equation}
E_{pol}=-\Sigma({\bf k},\xi_{\bf k})|_{{\bf k}=0},
\end{equation}
where $\xi_{\bf k}=\hbar^2 k^2/2m_{b}$ with $m_{b}$ being 
the band mass is the parabolic band dispersion. The 
nonparabolicity of the energy dispersion is ignored.  
The effective mass in the same method is defined as
\begin{equation}
\frac{m_{b}}{m^{\ast}}=1+\biggl(\frac{m_{b}}{\hbar^2}\biggr)
\frac{\partial^{2}\Sigma({\bf k},\xi_{\bf k})}{\partial k^2}\Biggr|_{{\bf k}=0},
\end{equation}

The electron-electron and electron-phonon interactions in the form
\begin{equation}
V_{d}(q,\omega)=\frac{v_{d}(q)}{\epsilon_{d}(q,\omega)}+
\frac{M_{d}(q)^2}{\epsilon_{d}(q,\omega)^2}
D_{d}(q,\omega)
\end{equation}
where $v_{d}(q)$ is the Fourier transform of the Coulomb potential, 
$\epsilon_{d}(q,\omega)$ is the dielectric function, $|M_{d}(q)|^2$ is the 
electron-phonon interaction term and $D_{d}(q,\omega)$ is the phonon 
renormalized Green's function. 

The strength of electron-phonon interaction $M_d(q)$
of a material depenends on the optical properties 
of that material. In an anisotropic 
low dimensional structure, there are confined, 
half-space and interface phonon modes
as a consequence of the presence of heterointerfaces. 
Consequently, a rigorous treatment 
of the electron-phonon interaction in semiconductor heterostructures 
requires the consideration
of all these modes\cite{Hai1,Hi2}. Since the anisotropic Euclidean space in low
dimensional structures are treated isotropic in FD space, 
the electron phonon interaction is treated similar to bulk modes. 

The renormalized phonon propagator is defined as
\begin{equation}
D_{d}(q,\omega)=\frac{2\omega_{LO}}
{\omega^2-\omega^{2}_{LO}-|M_{d}(q)|^2
\omega_{LO}\chi^{0}_{d}(q,\omega)/\epsilon_{d}(q,\omega)},
\end{equation}
where $\chi^{0}(q,\omega)$ is the irreducible polarizability function. 
In the single-pole plasmon approximation, $\chi^{0}(q,\omega)$ is defined as 
\begin{equation}
\chi^{0}_{d}(q,\omega)=\frac{nq^2}{m_{b}}\frac{1}
{\omega^{2}-\Omega^{2}_{d}(q)+\omega^2_{d}(q)}
\end{equation}
In the PPA method $\epsilon(q,\omega$ is expressed as
\begin{equation}
\frac{1}{\epsilon_{d}(q,\omega)}=\frac{1}{\epsilon_{\infty}}
\Biggl[1+\frac{\omega^2_{d}(q)}
{\omega^{2}-\Omega^{2}_{d}(q)}\Biggr]
\end{equation}
The plasmon frequency in the long-wavelength limit within 
the PPA method is given as\cite{Panda}
\begin{equation}
\Omega^{2}_{d}(q)=\omega^{2}_{d}(q)+q^2v_{F}^2\Biggl[\frac{1}{d}-
\frac{(7-d)\Gamma\small(\frac{d}{2}\small)}
{4(d+2)\pi^{\frac{1}{2}}\Gamma(\frac{d+3}{2}\small)}
\biggl(\frac{q_{TF}}{k_{F}}\biggr)^{d-1}\Biggr]
\end{equation}
In this equation $v_{F}$ is the Fermi velocity, $v_{F}=\hbar k_{F}/m^{\ast}$ 
with $k_{F}$ being the Fermi momentum. 
The Thomas-Fermi momentum is defined as
\begin{equation}
q^{d-1}_{TF}=\frac{2m_{b}e^2k^{d-2}_{F}}
{\sqrt{\pi}\epsilon_{\infty}\hbar^2}
\Biggl[1+\biggl\{\Gamma\biggl(1+
\frac{d}{2}\biggr)\biggr\}^2\Biggr]^{\frac{1}{d}}
\end{equation}
Using Eq.() and (), $D(q,\omega)$ is defined as
\begin{equation}
D_{d}(q,\omega)=\frac{2\omega_{LO}[\omega^2-\Omega^{2}_{d}(q)]}
{\hbar[\omega^2-\omega^{2}_{+}(q)][\omega^2-\omega^2_{-}(q)]}
\end{equation}
where
\begin{equation}
\omega^{2}_{\pm}(q)=\frac{1}{2}[\omega^{2}_{LO}+\Omega^2_{d}(q)]
\pm\frac{1}{2}\sqrt{[\omega^{2}_{LO}-\Omega^{2}_{d}(q)]^{2}+
4\omega^{2}_{d}(q)(\omega^{2}_{LO}-\omega^{2}_{TO})}
\end{equation}
In the presence of free electrons the LO-phonon frequency is renormalized 
due to plasmon-phonon coupling giving rise to two coupled modes with 
frequencies $\omega_{+}(q)$ and $\omega_{-}(q)$ which lie close to the 
uncoupled modes away from the mode-coupling region.  

The leading-order contribution to the electron self-energy due to the 
electron-phonon interaction is given by
\begin{equation}
\Sigma_{d}({\bf k},ik_{m})=-\frac{1}{\beta}
\sum_{{\bf q}}\sum_{i\omega_{n}}\frac{|M(q)|^2}{\epsilon(q,i\omega_{n})^2}
D(q,i\omega_{n})G^{0}({\bf k}+{\bf q},ik_{m}+i\hbar\omega_{n}),
\end{equation}

where $G^{0}({\bf k}+{\bf q},ik_{m}+i\hbar\omega_{n})$ is the 
free electron Green function and  
$ip_{m}$ and $i\omega_{n}$ are the 
standard Fermi and Bose imaginary frequencies,respectively, in the Matsubara's 
formalism. 
The free electron Green function $G^{0}(\iota\omega_{n},\xi_{(p+q)})$
is given as
$$G^{0}(k+q,\iota k_m+\iota\omega_{n})=\frac{1}
{\iota k_{m}+\iota\hbar\omega_{n}-\xi_{k+q}}$$

This is the one-phonon self-energy term which is justified
in the weak coupling limit. 
The purely electronic dielectric function 
$\epsilon(q,i\omega_{n})$ contains all informations about screening.    
Substituting Eqs(20), (23) and (26) in Eq.(25), we find
\begin{eqnarray}
\Sigma({\bf k},ik_{m})=-\frac{1}{\beta}\sum_{{\bf q}}|M(q)|^{2}
\sum^{\infty}_{i\omega_{n}=-\infty}&&
\frac{2\omega_{LO}[(i\omega_{n})^{2}-\Omega^{2}(q)+\omega^{2}_{p}(q)]^2}
{[(i\omega_{n})^2-\omega^{2}_{+}(q)][(\omega_{n})^2-\omega^{2}_{-}(q)]
[(i\omega_{n})^{2}-\Omega^{2}(q)]}\nonumber \\
&& \times\frac{1}{ik_{m}+i\hbar\omega_{n}-
\xi_{{\bf k}+{\bf q}}}
\end{eqnarray}
 
The classical method outlined in Ref.9 is followed to carry out 
the frequency summation. First of all we form the triplet 
$(\omega_{+}\omega_{-}\Omega)$ and indicate it as 
$(\omega_{1}\omega_{2}\omega_{3})$. The $i\omega_{n}$ summation 
is converted into an integral as $I=\int dzf(z)n_{b}(z)/2\pi i$ 
where $z=i\omega_{n}$ and $n_{B}$ is the Bose distribution function.
Using the contour integration method the poles and residues of $n_{B}(z)$
and $f(z)$ are determined. The integral $I$ is then found by addiding 
all residues. In the Jordan's lemma $I=0$ when $R\rightarrow\infty$.  
Finally in the analytic continumm $ip_{m}=\xi_{\bf k}+i\delta$, 
a final expression for the self-energy has been derived as  
\begin{eqnarray}
\Sigma({\bf k},\xi_{\bf k})=\sum_{\bf q}|M_{d}(q)|^2\sum_{ijk}\epsilon_{ijk}
&&\Biggl[\frac{F_{ijk}(q)
n_{B}(\omega_{i})
+G_{ijk}({\bf p},{\bf q})n_{F}(\xi_{{\bf p}+{\bf q}})}
{\xi_{\bf p}-\xi_{{\bf p}+{\bf q}}+\hbar\omega_{i}+i\delta}\nonumber \\
&& +\frac{F_{ijk}(q)[n_{B}(\omega_{i})+1]-
G_{ijk}({\bf p},{\bf q})n_{F}(\xi_{{\bf p}+{\bf q}})}
{\xi_{\bf p}-\xi_{{\bf p}+{\bf q}}-\hbar\omega_{i}+i\delta}\Biggr],
\end{eqnarray}
where $\epsilon_{ijk}$ is defined as 
$\epsilon_{123}=\epsilon_{231}=\epsilon_{312}=1$ and all other combinations 
$ijk$ are zero, $F_{ijk}(q)$ is obtained as 
\begin{equation}
F_{ijk}(q)=\frac{(\omega^2_{i}(q)-\Omega^2(q)+
\omega^{2}_{p}(q))^{2}
\omega_{LO}}{\omega_{i}(q)[\omega^2_{i}(q)-\omega^2_{j}(q)]
[\omega^2_{i}(q)-\omega^2_{k}(q)]}
\end{equation}
and $G_{ijk}({\bf k},{\bf q})$ is derived as
\begin{equation}
G_{ijk}({\bf k},{\bf q})=\frac{\omega_{LO}[(\xi_{{\bf k}+{\bf q}}-
\xi_{\bf k})^2-\hbar^2\Omega^2_{d}(q)
+\hbar^2\omega^{2}_{d}(q)]^2}
{\hbar\omega_{i}(q)[(\xi_{{\bf k}+{\bf q}}-
\xi_{\bf k})^2-\hbar^2\omega^{2}_{j}(q)]
[(\xi_{{\bf k}+{\bf q}}-\xi_{\bf k})^2-\hbar^2\omega^2_{k}(q)]}
\end{equation}

At zero-temperature, both $n_{F}$ and $n_{B}$ vanish and
the expression for the self-energy is obtained as
\begin{equation}
\Sigma({\bf k})=-\sum_{\bf q}|M_{d}(q)|^2
\Biggl[\frac{F_{123}(q)}
{\xi_{\bf k}-\xi_{{\bf k}+{\bf q}}-\hbar\omega_{+}(q)}+
\frac{F_{231}(q)}{\xi_{\bf k}-\xi_{{\bf k}+{\bf q}}-\hbar\omega_{-}(q)}+
\frac{F_{312}(q)}{\xi_{\bf k}-\xi_{{\bf k}+{\bf q}}-\hbar\Omega(q)}\Biggr]
\end{equation}
 
Using Eq.(\ref{eq:sum}) in Eq.(\ref{eq:self}) and then substituing 
Eq.(), we find

\begin{equation}
E_{p}=-\alpha C
\int dq\Biggl[\frac{F_{123}(q)}{q^2+2m_{b}\omega_{+}(q)/\hbar}+
\frac{F_{231}(q)}{q^2+2m_{b}\omega_{-}(q)/\hbar}+
\frac{F_{312}(q)}{q^2+2m_{b}\Omega(q)/\hbar}\Biggr]
\end{equation}
where $C=2m_{b}\omega^{2}_{LO}R_{p}\Gamma[(d-1)/2]/\sqrt{\pi}\Gamma[d/2]$.
The effective mass can be obtained as
Taking the coefficient of $k^2$ term, the effective mass is obtained as
\begin{equation}
\frac{m}{m^{\ast}}=1+\alpha\frac{8m_{b}C}{\hbar^2\sigma}
\int dq q^2\Biggl[\frac{F_{123}(q)}
{(q^2+2m_{b}\omega_{+}(q)/\hbar)^3}+
\frac{F_{231}(q)}{(q^2+2m_{b}\omega_{-}(q)/\hbar)^3}+
\frac{F_{312}(q)}{(q^2+2m_{b}\Omega(q)/\hbar)^3}\Biggr]
\end{equation}
\section{result and discussion}

The polaron binding energy and effective masses due to dynamic 
screening correction of electron-phonon interaction
is calculated for several $r_{s}$ values 
for dimensions $\beta$=2, 2.5 and 3 are shown in Table II. 
respectively. Here also the polaron properties 
 are found to increase with the increasing  $r_{s}$ and hense
decrease with the  increasing carrier density. 
This results due to the dynamic screening of the electron-phonon interaction
is slihgtly greater than that of static screening. 

The polaron properties due to dynamic screening also decrease as 
the dimension decreases which shows little higher values than 
static screening. 

Here we have taken same papametres as in static scrrening such as
 the band mass  $m_b$=0.067$m_{0}$, where $m_{0}$ 
is the electron mass,   
$\epsilon_{\infty}=13.18$ and $\epsilon_{0}=10.89$. The value of 
$\alpha$=0.067 which is appropriate for calculating polaron properties 
in the weak coupling limit. The LO phonon energy $(\hbar\omega_{LO})$ 
is taken as 36.25 meV.

\subsection{electron-electron interaction(plasmaron)}

The self energy equation for the electron-electron interaction is given by 
 is given by
$$\Sigma_d(p,\iota k_m)=-\frac{1}{\beta}\sum_{m,q}
V_{eff}G^0(\iota\omega n,\xi_{p+q})$$

Here $V_{eff}=\frac{V_o(q)}{\epsilon(q,\omega)}$ is the effective potential for
electron-electron interaction term where $V_o(q)$ is the Fourier 
transform of Coulomb potential and $\epsilon(q,\omega)$ is the screened 
dielectric function.

$$\Sigma_{i,j}(p)=-\frac{V_0(q)}{\epsilon_{\infty}}\sum_{m,q}\left[
\frac{n_F(\xi_{p+q})\left((\xi_{p+q}-
\iota\omega_n)^2+\tilde{\omega}_p^2-\omega_p^2\right)
-\tilde{\omega}_p^2(n_B(\omega_p)+1)}
{2\omega_p(\xi_{p}-\xi_{p+q}-\hbar\omega)}-
\frac{n_F(\xi_{p+q})\left((\xi_{p+q}-\iota\omega_n)^2+
\tilde{\omega}_p^2-\omega_p^2\right)+
\tilde{\omega}_p^2n_B(\omega_p)}
{2\omega_p(\xi_{p}-\xi_{p+q}-\hbar\omega)}\right]$$

At zero temperature both $n_F$ and $n_B$ vanish and the self energy 
equation becomes
$$\Sigma_{i,j}(p)=-\frac{V_0(q)}{\epsilon_{\infty}}\sum_q
\frac{\tilde{\omega}_p^2}{2\omega_p(\xi_{p}-\xi_{p+q}-\hbar\omega)}$$
The effective mass and binding energy equation are derived
from the above self energy equation proceeding with eq(1) and (2).

$$E_{p}=-\alpha C\int dq\Biggl[\frac{\tilde{\omega}_p^2}
{2\omega_p(q^2+2m_{b}\Omega(q)/\hbar)}\Biggr]$$
where $C=2m_{b}\omega^{2}_{LO}R_{p}\Gamma[(d-1)/2]/\sqrt{\pi}\Gamma[d/2]$.
The effective mass can be obtained as
Taking the coefficient of $k^2$ term, the effective mass is obtained as
$$\frac{m}{m^{\ast}}=1+\alpha\frac{8m_{b}C}{\hbar^2\sigma}
\int dq q^2\Biggl[\frac{\tilde{\omega}_p^2}
{2\omega_p(q^2+2m_{b}\Omega(q)/\hbar)^3}\Biggr]$$

\section{Results and  Discussions}

The polaron binding energy and effective masses due to dynamic 
screening correction of electron-electron  interaction
is calculated for several $r_{s}$ values 
for dimensions $\beta$=2, 2.5 and 3 are shown in Table III.
Here also binding  energies and effective masses are found to increase with 
the increasing  $r_{s}$ and hense 
decrease with the  increasing carrier density. 
The polaron properties also decrease with increase in dimension.

Here we have taken same papametres as in previous such as
 the band mass  $m_b$=0.067$m_{0}$, where $m_{0}$ 
is the electron mass,   
$\epsilon_{\infty}=13.18$ and $\epsilon_{0}=10.89$. The value of 
$\alpha$=0.067 which is appropriate for calculating polaron properties 
in the weak coupling limit.

\begin{figure}[htb]
\includegraphics[width=1.1\textwidth]{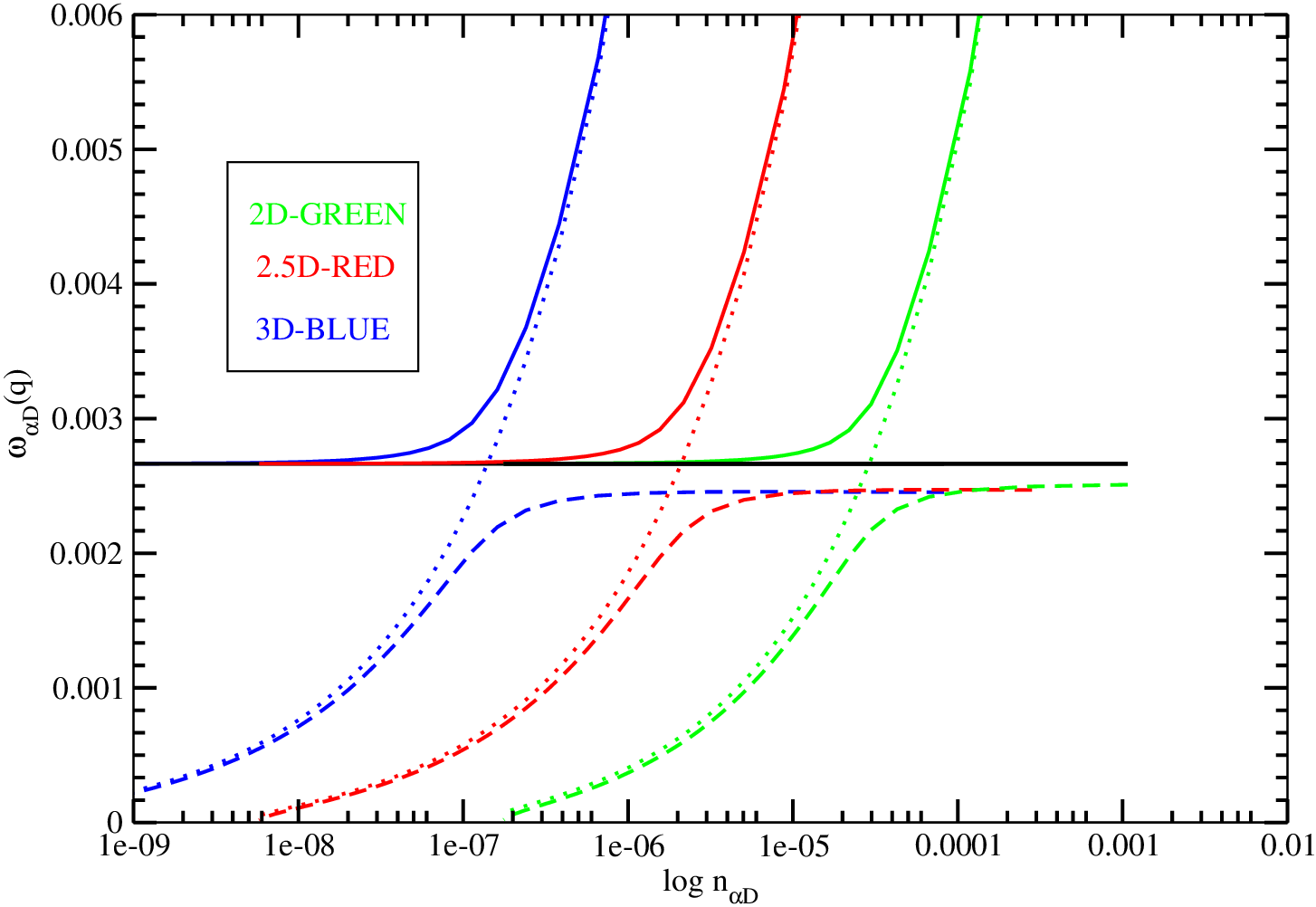}
\caption{Frequency of plasma modes as a function of logarithimic 
carrier density$n_{D}$ in fractional dimension. $\omega_+$(solid line),
$\omega_-$(dashed line) and $\omega_p$(dotted line) are shown for 2D, 
2.5D and 3D in green,red and blue colour respectively. The black line 
shows the LO phonon frequency.} 
\label{fig:class}
\end{figure}

\begin{figure}[htb]
\includegraphics[width=1.1\textwidth]{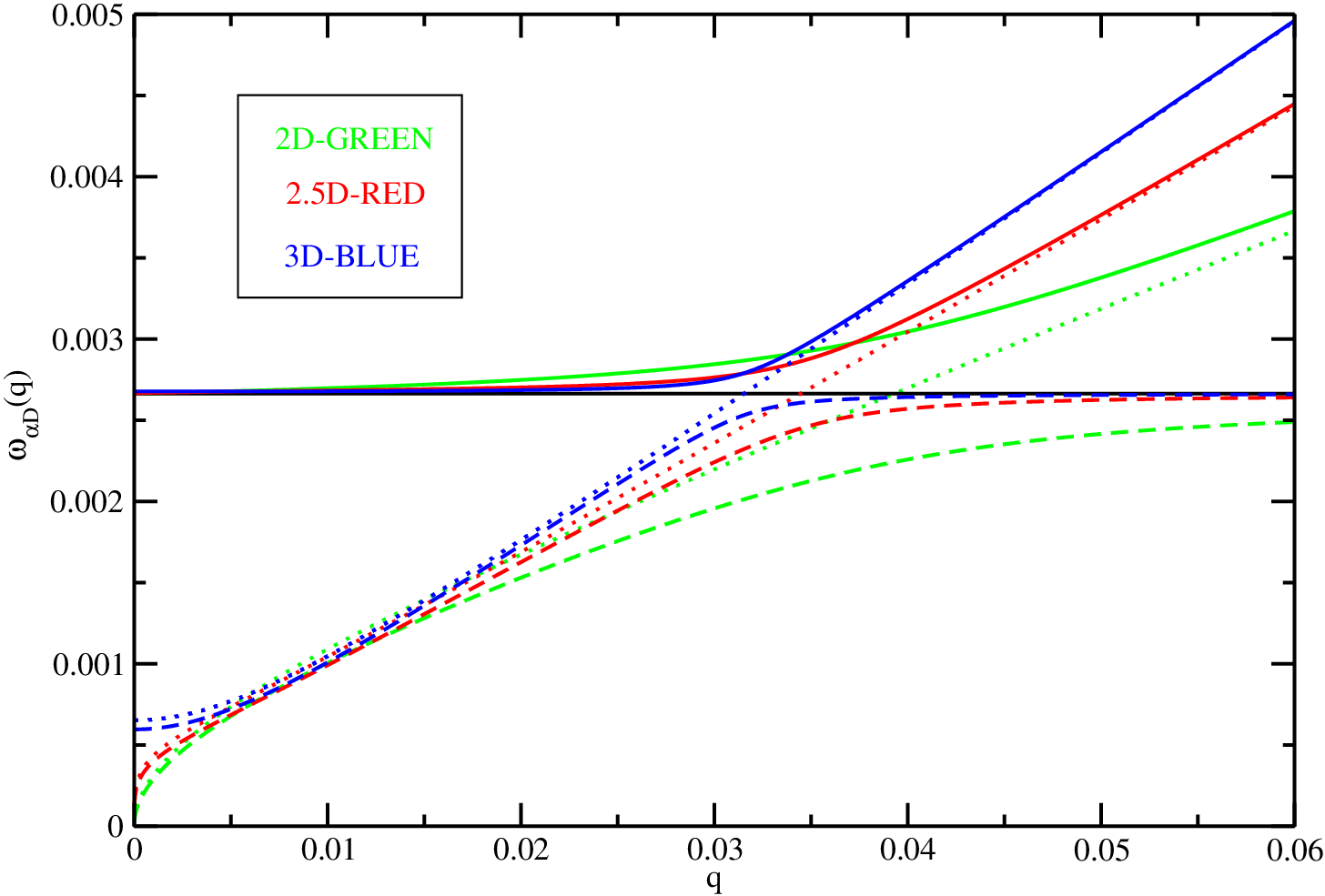}
\caption{Frequency of plasma modes as a function of wave vector q in fractional dimension. $\omega_+$(solid line),
$\omega_-$(dashed line) and $\omega_p$(dotted line) are shown for 2D, 
2.5D and 3D in green,red and blue colour respectively. The black line 
shows the LO phonon frequency.} 
\label{fig:class}
\end{figure}

\begin{figure}[htb]
\includegraphics[width=1.1\textwidth]{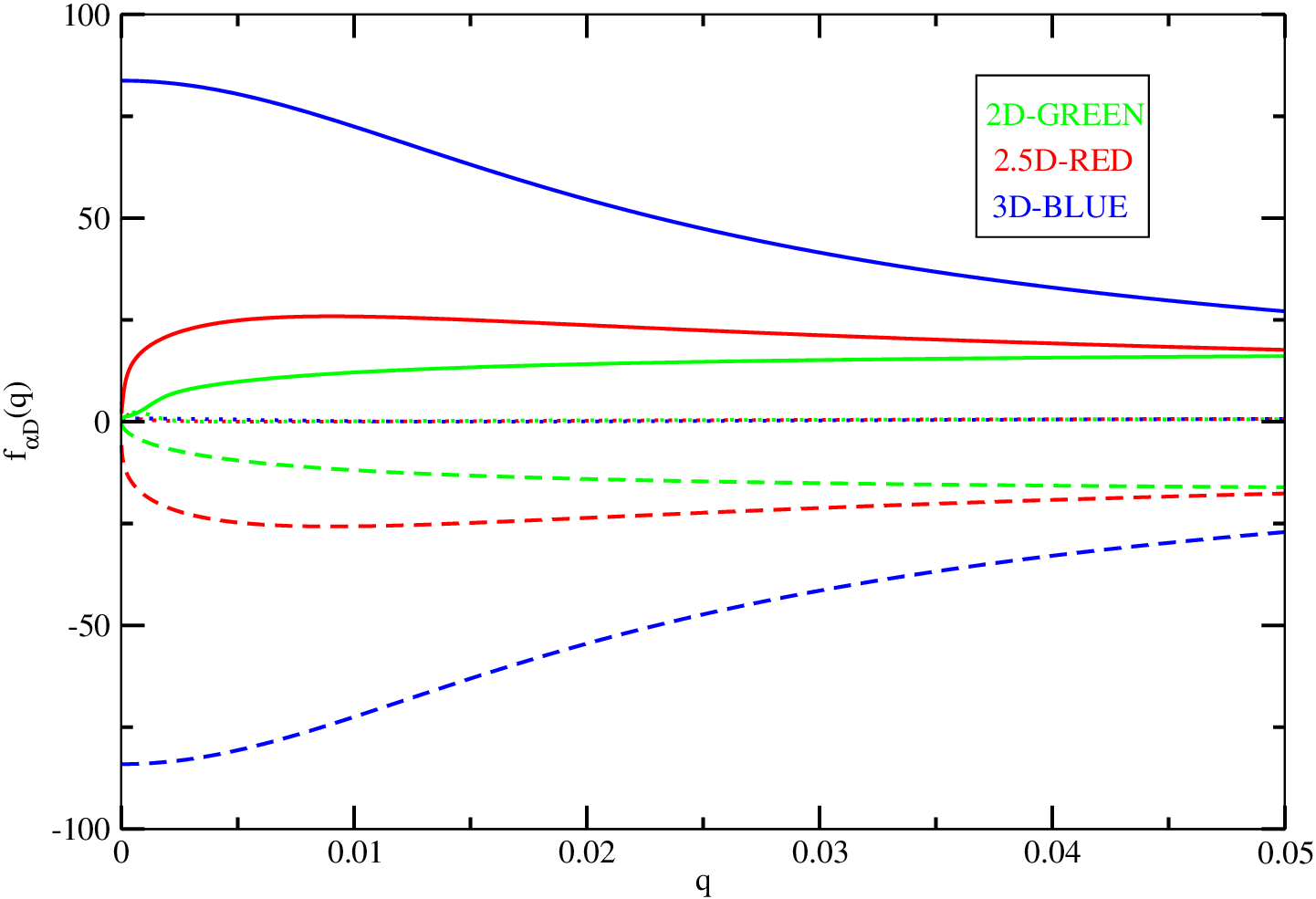}
\caption{The contribution factor F(q) as a function of wave vector q in 
fractional dimension.}  $F_+$(solid line),
$F_-$(dashed line) and $F_p$(dotted line) are shown for 2D, 
2.5D and 3D in green,red and blue colour respectively.  
\label{fig:class}
\end{figure}

\begin{table}
\begin{ruledtabular}
\begin{tabular}{ccccccc}
&\multicolumn{3}{c}{Polaron Energy (eV)}&\multicolumn{3}{c}{Effective mass}\\
\hline
$r_{s}$ & $\beta=2$ & $\beta=2.5$  & $\beta=3$ & $\beta=2$ & $\beta=2.5$ & 
$\beta$=3 \\
0.001 & 0.416 & 0.296 & 0.210 & 1.120 & 1.067 & 1.041 \\
0.01 & 0.417 & 0.315 & 0.255 & 1.121 & 1.070 & 1.047 \\
0.10 & 0.424 & 0.325 & 0.272 & 1.123 & 1.071 & 1.049 \\
1.0 & 0.430 & 0.328 & 0.274 & 1.124 & 1.072 & 1.050 \\
10.0 & 0.430 & 0.340 & 0.275 & 1.125 & 1.073 & 1.051 \\
\end{tabular}
\end{ruledtabular}
\end{table}

\begin{table}
\begin{ruledtabular}
\begin{tabular}{ccccccc}
&\multicolumn{3}{c}{Polaron Energy (eV)}&\multicolumn{3}{c}{Effective mass}\\
\hline
$r_{s}$ & $\beta=2$ & $\beta=2.5$  & $\beta=3$ & $\beta=2$ & $\beta=2.5$ & 
$\beta$=3 \\
0.001 & 0.424 & 0.081 & 0.011 & 1.177 & 1.009 & 1.000 \\
0.01 & 0.489 & 0.184 & 0.066 & 1.219 & 1.041 & 1.004 \\
0.10 & 0.554 & 0.356 & 0.221 & 1.260 & 1.136 & 1.058 \\
1.0 & 0.609 & 0.522 & 0.426 & 1.275 & 1.207 & 1.175 \\
10.0 & 0.656 & 0.613 & 0.538 & 1.280 & 1.210 & 1.178 \\
\end{tabular}
\end{ruledtabular}
\end{table}

\begin{table}
\begin{ruledtabular}
\begin{tabular}{ccccccc}
&\multicolumn{3}{c}{Polaron Energy (eV)}&\multicolumn{3}{c}{Effective mass}\\
\hline
$r_{s}$ & $\beta=2$ & $\beta=2.5$  & $\beta=3$ & $\beta=2$ & $\beta=2.5$ & 
$\beta$=3 \\
0.01 & 0.019 & 0.002 & 0.001 & 1.000 & 1.000 & 1.000 \\
0.05 & 0.094 & 0.011 & 0.004 & 1.010 & 1.005 & 1.003 \\
0.10 & 0.189 & 0.024 & 0.008 & 1.075 & 1.041 & 1.020 \\
0.5d0 & 0.934 & 0.139 & 0.037 & 6.263 & 3.970 & 3.067 \\
1.0 & 1.858 & 0.304 & 0.067 & 26.968 & 13.995 & 9.801 \\
\end{tabular}
\end{ruledtabular}
\end{table}

\end{document}